\documentstyle[11pt,psfig]{article}
\begin{document}
\title{\huge{Charmonia Production in Hadron Colliders and the 
Extraction of Colour-Octet Matrix Elements}
\thanks{Research partially supported by CICYT under grant AEN-96/1718}} 
\author{{\bf B. Cano-Coloma and M.A. Sanchis-Lozano
\thanks{Corresponding author}$\ $\thanks{ 
E-mail:mas@evalvx.ific.uv.es ; Phone: +34 6 386 4752 ; 
Fax  : +34 6 386 4583}
}
\\ \\
\it Instituto de F\'{\i}sica
 Corpuscular (IFIC), Centro Mixto Universidad de Valencia-CSIC \\
\it and \\
\it Departamento de F\'{\i}sica Te\'orica \\
\it Dr. Moliner 50, E-46100 Burjassot, Valencia (Spain) }
\maketitle 
\begin{abstract}
  We present the results of our analysis on charmonia ($J/\psi$ and ${\psi}'$) 
  hadroproduction taking into account higher-order QCD effects induced by 
  initial-state radiation in a Monte Carlo framework, with the colour-octet
  mechanism implemented in the event generation. We find that those 
  colour-octet matrix elements extracted so far from Fermilab Tevatron data
  for both $J/\psi$ and ${\psi}'$ production have to be lowered significantly. 
  We finally make predictions for charmonia production at the LHC, presenting
  a simple code for a {\em fast} simulation with PYTHIA based on the 
  colour-octet model.
\end{abstract}
\vspace{-13.5cm}
\large{
\begin{flushright}
  IFIC/97-29\\
  FTUV/97-30\\
  \today
\end{flushright} }
\vspace{13.5cm}
\newpage
\section{Introduction}
Likely, the Large Hadron Collider or LHC will become operative by the 
beginning of the next century, offering a wide programme of exciting 
possibilities in Particle Physics, among which the origin of mass at the 
electroweak scale and the possible discovery of supersymmetric 
particles, the investigation of CP violation in B mesons and detailed 
studies of top quark physics. Other interesting topics related to 
charm and beauty flavours will be covered as well benefitting of
a foreseen huge statistics to be collected with the machine running even
at $\lq\lq$low" luminosity (${\simeq}\ 10^{33}$ cm$^2s^{-1}$) \cite{atlas}
\cite{cms}. Following this line of research, we shall focus in this
paper on inclusive hadroproduction of heavy resonances, in particular 
charmonia $J/\psi$ and ${\psi}'$ states, whose relevance in the study of
perturbative and non-perturbative QCD will be briefly \vspace{0.1in} reviewed.
\par
To first approximation, heavy-flavour hadroproduction can be treated
in the framework of perturbative QCD due to the relatively large quark
masses \cite{ellis} and consequently the partonic production cross 
sections can be convoluted 
with parton distribution functions (PDF's) of the colliding protons.
This scheme can indeed be viewed as well founded since the 
underlying parton interaction is hard, therefore
providing a reasonable justification for such a factorization of the
production process. Higher-order corrections to the matrix elements
of the short-distance processes may be incorporated likely improving the 
accuracy of the predicted cross sections. Even complex effects such as 
multiple gluon emission can be treated analytically according to resummation 
techniques \vspace{0.1in} \cite{resum}. 
\par
Nevertheless, such an essentially analytic approach may eventually 
oversight the real complexity of the full hadronic collision and 
a certain amount of modifications should be added if a more realistic 
description is required. For example, parton evolution either at the initial
or final state together with fragmentation and recombination of the
beam remnants could play an important r\^{o}le leading to 
substantial modifications in the final state, even for
inclusive \vspace{0.1in} processes.
\par
On the other hand, event generators based on a Monte Carlo framework 
represent an alternative philosophy of coping with 
the complexity of very high energy hadronic reactions by dividing 
the full problem in a set of components to be dealt separately in a 
simulation chain. The hard partonic interaction, constituting the 
underlying or skeleton process of the simulation, is 
dressed up with soft processes leading to
hundreds of final particles. Certainly, such an approach  
is plagued with various sources of uncertainty, 
especially related to the long-distance aspects of the strong interaction
dynamics, associated to the assumptions and modelling of the generation. 
However, the impressive amount of collected
experimental data used in a heuristic way
to {\em tune} those parameters of the event generator, gives
reliability to this method to a considerable degree providing
a powerful tool to interpret current data and make sensible predictions 
for future \vspace{0.1in} experiments.
\par
In this work we re-examine prompt production of heavy resonances
at high-energy hadron colliders by means of the event generator PYTHIA
\cite{pythia}. Undoubtedly, there is a widespread 
recognition that the study of heavy quarkonia has been playing 
an important r\^{o}le over the past decades
in the development of the theory of the strong interaction and
such beneficial influence seems far from ending. Moreover, a precise 
understanding of prompt charmonia production offer several advantages 
from both experimental and theoretical points of view at the 
\vspace{0.1in} LHC. 
\par
The fact that both $J/\psi$ and ${\psi}'$ resonances can decay
into a oppositely charged lepton pair with a respectable branching
fraction (especially the former), permits its discrimination in a 
huge hadronic background. Actually, charmonia production can be 
viewed as interesting in its own right for precise tests of both 
perturbative and non-perturbative
QCD, search for exotic states or as a probe of the parton content
of the proton. Furthermore, completely different processes as
for example the decay $B_d^0{\rightarrow}{J/\psi}K_s^0$, a golden 
channel to investigate CP violation in B decays, require a precise
estimate of sources of background coming from random
combinations involving prompt \vspace{0.1in} $J/\psi$'s.
\par

On the other hand, the experimental
discovery at Fermilab \cite{fermi} of an excess of inclusive
production  of prompt heavy  quarkonia (mainly  for  $J/\psi$ and  $\psi'$
resonances) in  antiproton-proton collisions triggered an intense theoretical
activity beyond what was considered  conventional wisdom until recently
\cite{greco} \cite{beneke}. The discrepancy between the so-called colour-singlet
model (CSM) in hadroproduction \cite{baier,schuler0} and the experimental 
data amounts to more than an order of magnitude and cannot be attributed 
to those theoretical uncertainties arising from the ambiguities on the
choice of a particular parton distribution function, the heavy quark mass
or different energy \vspace{0.1in} scales.
\par 
The colour-singlet model assumes that the heavy quark pair is produced
in a colour-singlet state with the right quantum numbers already
during the hard collision. The production cross section involves
a perturbative calculation where the proper spin and angular
momentum states are projected out, and the non-perturbative
component of the hadronization comes from a single parameter to be
identified with the wave function of the
resonance at the origin (or its derivatives for non $S$-wave states). 
The fact that the latter can be derived from potential models fitting
the charmonia spectrum or from their leptonic decay modes,  
gives its predictive power to the colour-singlet model. However, as
mentioned above, such predictions stand quite below the
experimental yield at high-$p_t$, so that some modifications have 
to be done. In addition, it should be emphasized that the model
is clearly incomplete since relativistic corrections due to the
relative velocity of quarks in the bound state are ignored - 
expected to be sizeable especially for charmonia. Moreover, 
another hint of the incompleteness of the CSM comes from the
existence of infrared divergences in the production cross section 
for $P$-wave states, which cannot be absorbed in its non-perturbative
\vspace{0.1in} factor.
\par
In order to resolve all these difficulties, it has 
been recently argued that the heavy quark pair not necessarily has to be
produced in a colour-singlet state at the short-distance partonic process
itself \cite{braaten}. Alternatively, it can  be produced in a colour-octet
state evolving non-perturbatively into quarkonium in a specific final state
with the correct quantum numbers according to some computable  probabilities
governed by the internal velocity of the heavy quark. This mechanism, usually
named as  the colour-octet  model (COM)
\footnote{For the sake of brevity in our theoretical introduction 
we do not expose another approach to heavy quarkonia production
in the same spirit as the COM but based on duality arguments known as 
the colour-evaporation model, 
referring the reader to \cite{evap} and references therein}, can be cast 
into the rigorous framework of an effective non-relativistic theory for the 
strong interactions (NRQCD) deriving from first principles 
\vspace{0.1in} \cite{bodwin}.
\par

However, the weakness of the COM lies  in the fact that the non-perturbative
parameters characterizing the long-distance hadronization process beyond the
colour-singlet contribution (i.e. the colour-octet matrix elements)  are so
far almost free parameters to be adjusted  from the fit to experimental data,
though expected to be mutually consistent according to the  NRQCD power
scaling rules, or to resort to non-perturbative models 
\vspace{0.1in} \cite{schuler}.
\par 

On  the other  hand, an  attractive feature of the colour-octet hypothesis
consists of the  universality of the NRQCD matrix elements (ME) entering in 
other charmonium production processes like photoproduction \cite{hera}.  Let 
us look below in some detail at the way
hadronization is  folded with the partonic  description of  hadrons, focusing
for concreteness on the couple of related papers \vspace{0.1in} 
\cite{cho0,cho}. 
\par

In the first paper \cite{cho0}, Cho and Leibovich considered for charmonium
production (in addition to the CSM) the $^3S_1$ coloured intermediate state as 
a first approach, computing the matrix elements as products of perturbative
parts for the short-distance partonic  processes, and the colour-octet 
ME concerning the long-distance hadronization. Finally, a
convolution of concrete  parton distribution functions and the differential 
cross-section for the $Q\bar{Q}$  production subprocess was performed, whereby
the $p_t$ dependence of the charmonia production exclusively coming from 
the \vspace{0.1in} latter.
\par 
In Ref. \cite{cho} the same authors take into account further contributions 
from new coloured states (for more details see the quoted references) 
concluding finally that at high enough $p_t$ a two-parameter fit is 
actually required to explain the observed inclusive $p_t$ distribution of 
charmonia production at the Fermilab \vspace{0.1in} Tevatron.
\par

Notice, however, that such calculations were carried out based on a possibly
oversimplified picture of the hadronic interaction. Indeed, it is  
well-known for a long time that higher-order effects ($K$
factors) play an important r\^{o}le in inclusive hadroproduction. In 
particular, beyond the primordial transverse  momentum $k_t$ of
partons in hadrons related to Fermi motion relevant at small $p_t$,
initial-state radiation of gluons  by the interacting partons add up to 
yield an  {\em effective} intrinsic transverse momentum which certainly has 
to be considered in hadroproduction at high $p_t$ \cite{break}. As we 
shall see, if overlooked at all, the effect on the fit parameters 
(and ultimately on  the colour-octet ME's) amounts to a {\em systematic}
\vspace{0.1in} overestimate.
\par
In fact, one should distinguish the
primordial transverse momentum of partons owing to their Fermi motion
(a non-perturbative effect) from the perturbative contribution dynamically 
generated via gluon radiation generally implemented in the event generators
by means of a parton shower algorithm \cite{sjos} \footnote{However, it 
is not yet completely clear how to separate uniquely both components from real
experimental data \cite{break}}. By {\em effective} $k_t$ we merge both
effects under a common name, though the former is actually overshadowed by 
the latter at high $p_t$. Of course, the PYTHIA treatment of the 
effective $k_t$ is not guaranteed to be perfect but, nevertheless, should
give a reasonable estimate of such \vspace{0.1in} effects.
\par 
In this work we have extended our earlier analysis of $J/\psi$ 
hadroproduction at colliders \cite{bea} \cite{plb} to the ${\psi}'$ 
state, affording a more complete test upon the validity of the NRQCD 
power scaling rules relating different long-distance matrix elements. 
The paper is organized as follows. In Section 2 we describe the hard 
processes considered for charmonia hadroproduction and we discuss their
implementation in the event generator PYTHIA, described in more detail 
in the Appendix at the end of the paper where we also present a code
for a fast simulation of charmonia using PYTHIA. In Section 3 we perform
new fits to Tevatron data on prompt $J/\psi$ and ${\psi}'$ direct
production extracting NRQCD matrix elements from them. We finally
make some predictions for charmonia direct production at the 
LHC in Section \vspace{0.1in} 4.
\par

\section{Implementation of the COM in PYTHIA}

Originally, the event generator PYTHIA \cite{pythia} produces
direct $J/\psi$'s and higher ${\chi}_{cJ}$ resonances in
hadron-hadron collisions according to the CSM. However, as already mentioned
this model fails off to account for the charmonia production rate
at the Fermilab Tevatron by more than an order of magnitude. Consequently, 
we have implemented a code in PYTHIA in order
to include the colour-octet mechanism for $J/\psi$ hadroproduction
($\psi'$ generation has been based upon the same set of production
channels) via the following ${\alpha}_s^3$ partonic \vspace{0.1in} processes:
\begin{itemize}
\item $g\ +\ g\ {\rightarrow}\ J/\psi\ +\ g$ 
\item $g\ +\ q\ {\rightarrow}\ J/\psi\ +\ q$ 
\item $q\ +\ \bar{q}\ {\rightarrow}\ J/\psi\ +\ g$
\end{itemize}
\vspace{0.1in}
\par
We have included in the simulation the $^3S_1^{(8)}$, $^1S_0^{(8)}$ and
$^3P_J^{(8)}$ ($J=0,1,2$) states in $^3S_1$ charmonia 
hadroproduction according to the COM \cite{cho}. The squared amplitudes, 
taken from Refs. \cite{cho0,cho} are reproduced below. Firstly let us 
consider the $^3S_1^{(8)}$ intermediate state:
\begin{eqnarray}
\overline{\sum}{\mid}{\cal A}(gg{\rightarrow}{J/\psi}g){\mid}^2 & = &
-\frac{8{\pi}^3{\alpha}_s^3}{9M^3}\ \frac{27(\hat{s}\hat{t}+
\hat{t}\hat{u}+\hat{u}\hat{s})-19M^4}
{[(\hat{s}-M^2)(\hat{t}-M^2)(\hat{u}-M^2)]^2} \nonumber \\
& {\times} & [(\hat{t}^2+\hat{t}\hat{u}+\hat{u}^2)^2-M^2(\hat{t}+\hat{u})
(2\hat{t}^2+\hat{t}\hat{u}+2\hat{u}^2)+M^4(\hat{t}^2+\hat{t}\hat{u}+\hat{u}^2)]
\nonumber \\
& {\times} & <0{\mid}O_8^{J/\psi}(^3S_1){\mid}0> \\ \nonumber \\ 
\overline{\sum}{\mid}{\cal A}(gq{\rightarrow}{J/\psi}q){\mid}^2 & = &
-\frac{16{\pi}^3{\alpha}_s^3}{27M^3}\ \frac{\hat{s}^2+\hat{u}^2+2M^2\hat{t}}
{\hat{s}\hat{u}(\hat{t}-M^2)^2}[4(\hat{s}^2+\hat{u}^2)-\hat{t}\hat{u}]\ 
<0{\mid}O_8^{J/\psi}(^3S_1){\mid}0> \nonumber \\
& & 	\\
\overline{\sum}{\mid}{\cal A}(q\overline{q}{\rightarrow}{J/\psi}g){\mid}^2 & = &
\frac{128{\pi}^3{\alpha}_s^3}{81M^3}\ 
\frac{\hat{t}^2+\hat{u}^2+2M^2\hat{s}}
{\hat{t}\hat{u}(\hat{s}-M^2)^2}[4(\hat{t}^2+\hat{u}^2)-\hat{t}\hat{u}]\ 
<0{\mid}O_8^{J/\psi}(^3S_1){\mid}0> \nonumber \\
& &
\end{eqnarray} 
where the barred summation symbol refers to average over initial
and final spins and colours; 
$\hat{s}$, $\hat{t}$ and $\hat{u}$ stand for the Mandelstam variables 
of the short-distance subprocesses. We shall set $M=2M_c$ for both $J/\psi$ 
and ${\psi}'$ resonances in accordance with the assumption
on the non-relativistic nature of heavy quarkonium made in the 
theoretical calculation of these amplitudes \cite{cho0,cho}, i.e. vanishing 
relative momentum of the quarks in the bound \vspace{0.1in} state. 
\par
The corresponding expressions for the $^1S_0^{(8)}$ contributions 
\vspace{0.1in} are:
\begin{eqnarray}
\overline{\sum}{\mid}{\cal A}(gg{\rightarrow}{J/\psi}g){\mid}^2 & = &
\frac{20{\pi}^3{\alpha}_s^3}{M}\ \frac{(\hat{s}^2-M^2\hat{s}+M^4)^2
-\hat{t}\hat{u}(2\hat{t}^2+3\hat{t}\hat{u}+2\hat{u}^2)}
{\hat{s}\hat{t}\hat{u}[(\hat{s}-M^2)(\hat{t}-M^2)(\hat{u}-M^2)]^2} 
\nonumber \\
& {\times} & 
[\hat{s}^2(\hat{s}-M^2)^2+\hat{s}\hat{t}\hat{u}(M^2-2\hat{s})+
(\hat{t}\hat{u})^2]\ <0{\mid}O_8^{J/\psi}(^1S_0){\mid}0> \nonumber \\
& & \\
\overline{\sum}{\mid}{\cal A}(gq{\rightarrow}{J/\psi}q){\mid}^2 & = &
-\frac{40{\pi}^3{\alpha}_s^3}{9M}\ \frac{\hat{s}^2+\hat{u}^2}
{\hat{t}(\hat{t}-M^2)^2}\ <0{\mid}O_8^{J/\psi}(^1S_0){\mid}0> \\
\overline{\sum}{\mid}{\cal A}(q\overline{q}{\rightarrow}{J/\psi}q){\mid}^2 & = &
\frac{320{\pi}^3{\alpha}_s^3}{27M}\ \frac{\hat{t}^2+\hat{u}^2}
{\hat{s}(\hat{s}-M^2)^2}\ <0{\mid}O_8^{J/\psi}(^1S_0){\mid}0> 
\end{eqnarray}
\vspace{0.1in}
\par
With regard to the $^3P_J^{(8)}$ contributions, they display 
altogether a similar (i.e. degenerate) transverse momentum behaviour as 
the $^1S_0^{(8)}$ component for $p_t\ {\ge}\ 5$ GeV \cite{cho}. Thus from 
a pragmatic point of view, the generation of charmonia via
intermediate $P$-wave coloured states becomes superfluous
although they must properly be taken into account in the computation of
the overall cross section as explained in more detail in the 
\vspace{0.1in} Appendix.
\par
We find from our simulation that gluon-gluon scattering actually stands 
for the dominant process as expected, the gluon-quark scattering 
contributes appreciably ${\simeq}\ 20\%$, whereas the 
quark-antiquark scattering represents a tiny contribution. Therefore we 
shall omit hereafter any further reference to the $q\overline{q}$ production 
\vspace{0.1in} channel. 
\par

\begin{figure}[htb]
\centerline{\hbox{
\psfig{figure=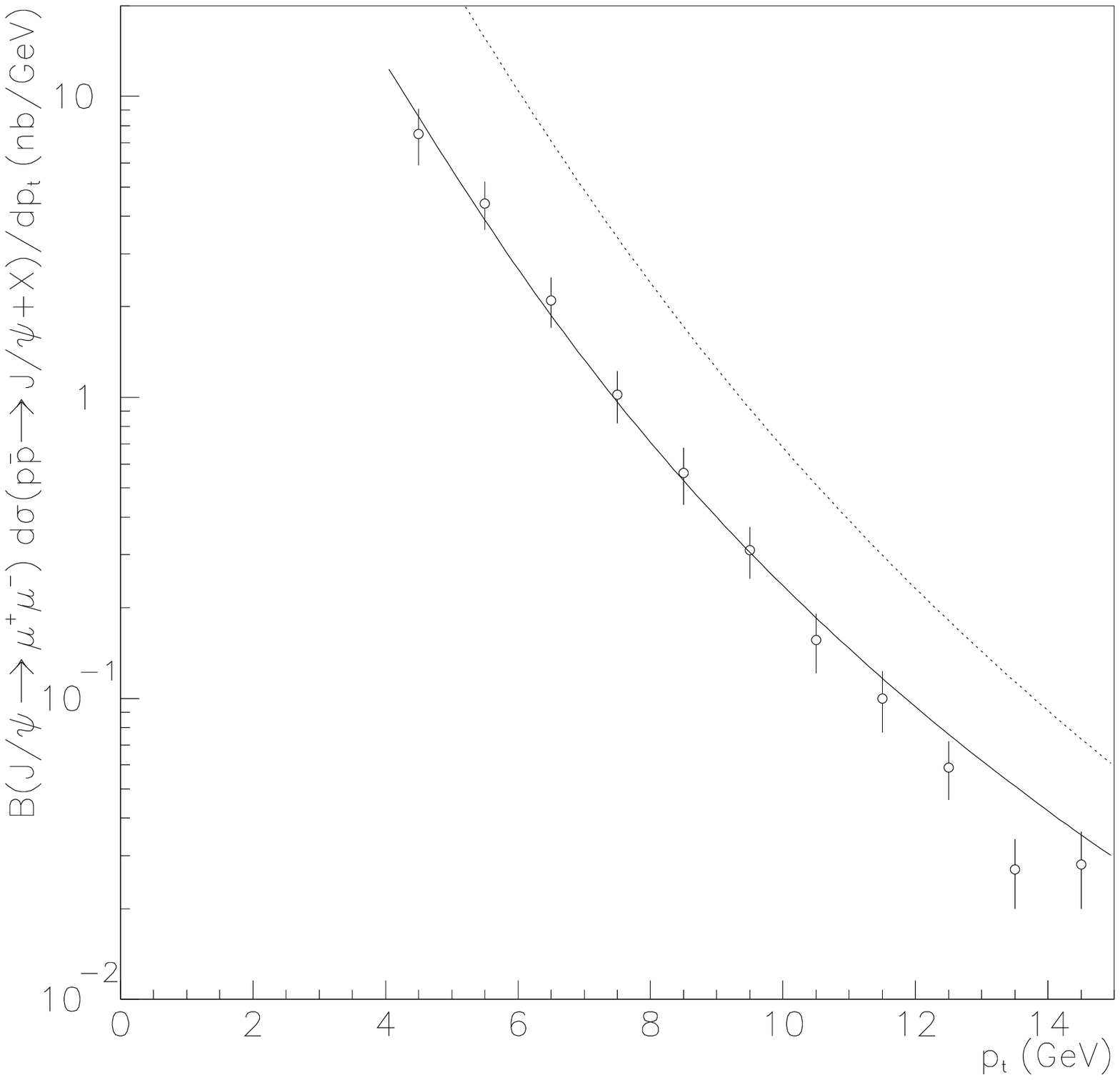,height=6.5cm,width=8.cm}
\psfig{figure=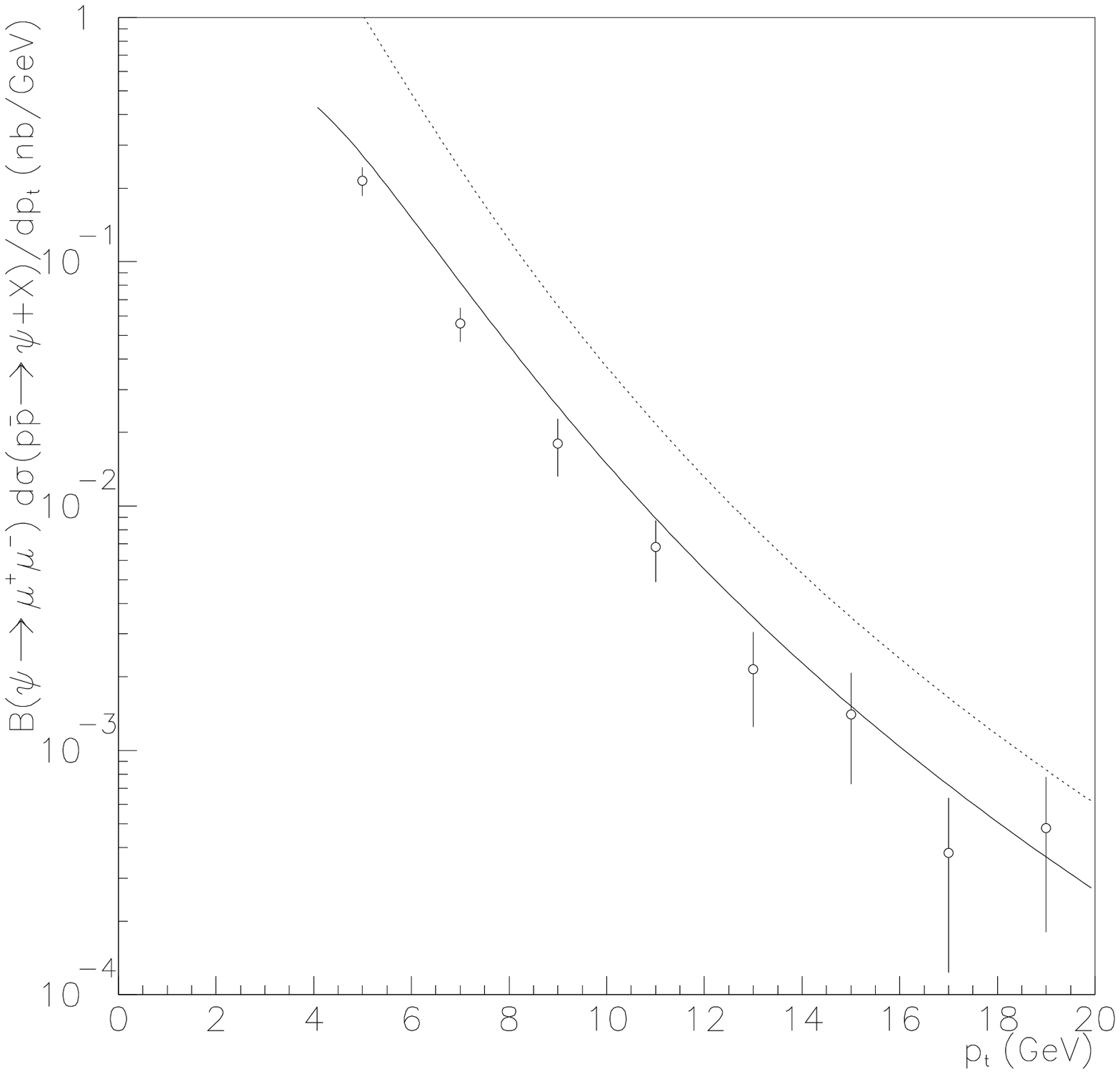,height=6.5cm,width=8.cm}
}}
\caption{ Curves obtained from PYTHIA (not fit) including the colour-octet
mechanism for prompt $J/\psi$ {\em (left)} and $\psi'$ {\em (right)} production 
at the Tevatron using the same parameters as in Ref. [17]. The solid line 
corresponds to initial- and final-state radiation off and the dotted 
line to initial- and final-state radiation on. The MRSD0 parton distribution
function and $M_c=1.48$ GeV were employed as in [17]. All curves
are multiplied by the corresponding muon branching fraction of charmonium.} 
\end{figure}

Using  the  same numerical  values  for the  colour-octet matrix  elements as
reported in Tables I and II of Ref. \cite{cho}, if initial- and final-state
radiation are  turned off, there is a good agreement between the
theoretical curve and the experimental points for both $J/\psi$ and
$\psi'$ (see Fig. 1), as should reasonably be expected . However, if
initial- and final-state radiation \footnote{It
should  be noted that  initial-state radiation and final-state radiation have
opposite effects in the $p_t$ spectrum, the former enhancing the high $p_t$
tail whereas the latter softens the distribution. Indeed, in considering
the process $gg{\rightarrow}{J/\psi}g$ in PYTHIA only the gluon evolves in the 
final state, though the energy (and momentum) of charmonium is modified as 
a consequence of the final-state machinery \cite{pythia}}  
are switched on, the predicted curve stands well above the experimental 
data, in accordance with the expected $k_t$-kick caused by the {\em
effective} intrinsic transverse momentum of partons \cite{break}. 
Accordingly, keeping radiation effects in the theoretical analysis 
it turns out that the values for the colour-octet ME's have to be 
lowered by significant \vspace{0.1in} factors. 

\par

For the sake of clarity it is important to point out that the
inclusion in the event generation of multi-gluon emission by
itself almost does not change the overall cross-section for
charmonia production obtained from PYTHIA. (There is however an
indirect bias effect slightly increasing the cross section
$\sigma({\mid}{\eta}_{J/\psi}{\mid}<0.6)$ with radiation on 
relative to radiation off because of the
kinematical cut on the charmonia pseudorapidity.) The main
effect remains in the smearing of the transverse momentum
distribution leading to the enhancement of the high-$p_t$
\vspace{0.1in} tail.

\section{Extraction of NRQCD matrix elements from Tevatron data}

In  order to assess the importance  of  the effective intrinsic
transverse momentum  of partons we have made  three different choices for the
proton PDF \footnote{See \cite{pdflib} for  technical  details about   the
package of  Parton Density Functions  available at the CERN Program Library.
References therein}: 

\begin{description} 
 \item[a)] the leading order CTEQ 2L (by default in PYTHIA 5.7)  
 \item[b)] the next to leading order MRSD0 (the same as used in \cite{cho}) 
 \item[c)] the next to leading order GRV 94 HO 
\end{description}

As mentioned  before, the  theoretical  curves for the  inclusive $p_t$
distribution of prompt $J/\psi$ and ${\psi}'$  stand in all cases above   
Tevatron experimental points if the set of parameters from \cite{cho} are 
blindly employed in the PYTHIA generation with initial-state radiation on.
Motivated by this systematic discrepancy, we have performed new fits for 
prompt $J/\psi$ and ${\psi}'$ direct production at the Tevatron 
(feed-down to $J/\psi$ from radiative decay of ${\chi}_{cJ}$ resonances 
was experimentally \vspace{0.1in} removed).
\par
Those colour-octet matrix elements extracted from our
analysis on $J/\psi$ production are reported in Table 1. In Fig. 2a
we plot the resulting fit to Tevatron data for the
CTEQ PDF, showing separately the three different components taken into
account in this work. As a figure of merit for the overall fit we
found ${\chi}^2/N_{DF}=1.8$. In the case of $\psi'$ production, the 
respective ME's are reported in Table 2 and the corresponding plot for 
the CTEQ PDF is shown in figure 2b, with \vspace{0.1in} ${\chi}^2/N_{DF}=0.4$. 
\par

\par 
\begin{figure}[htb]
\centerline{\hbox{
\psfig{figure=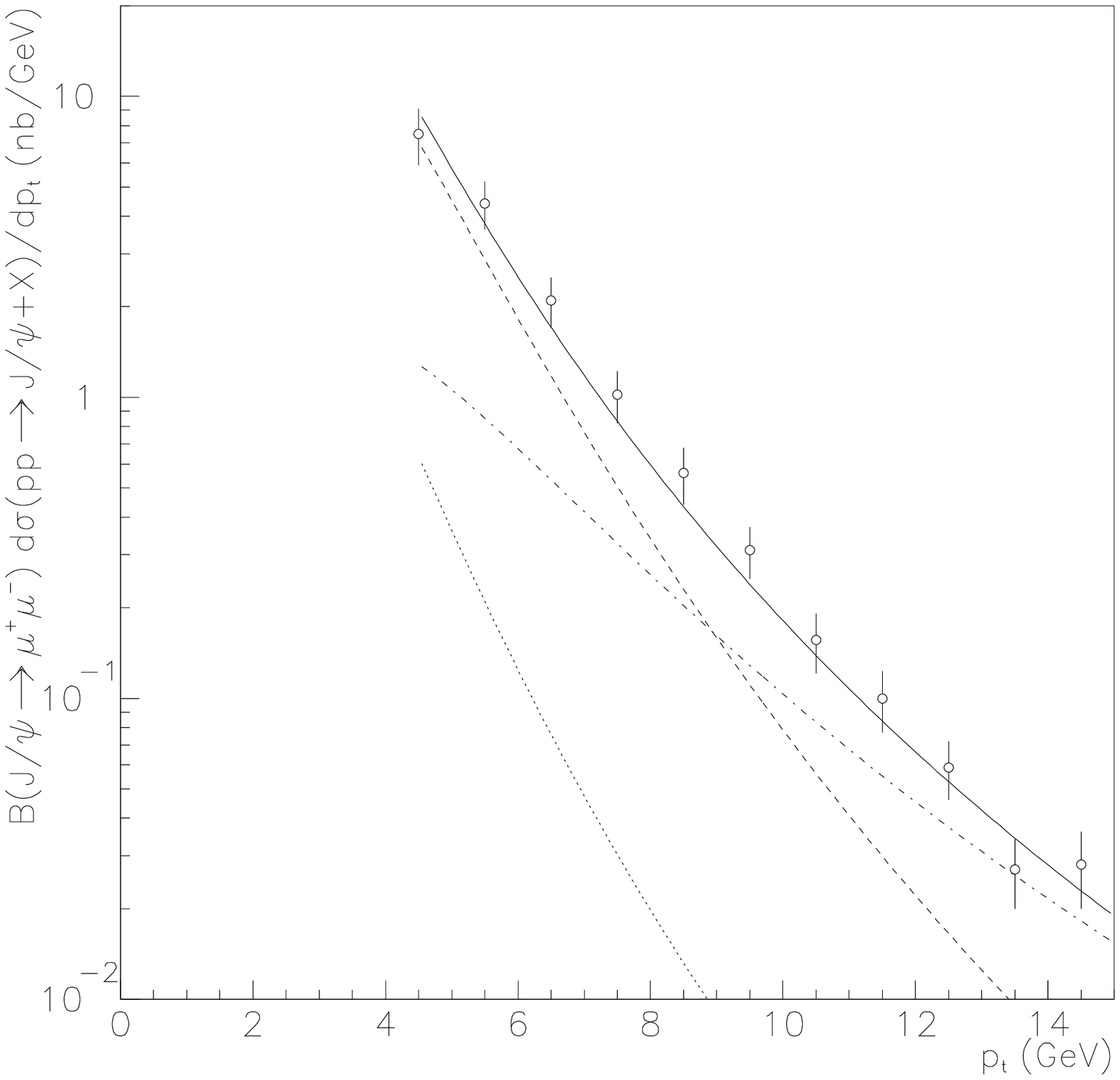,height=6.5cm,width=8cm}
\psfig{figure=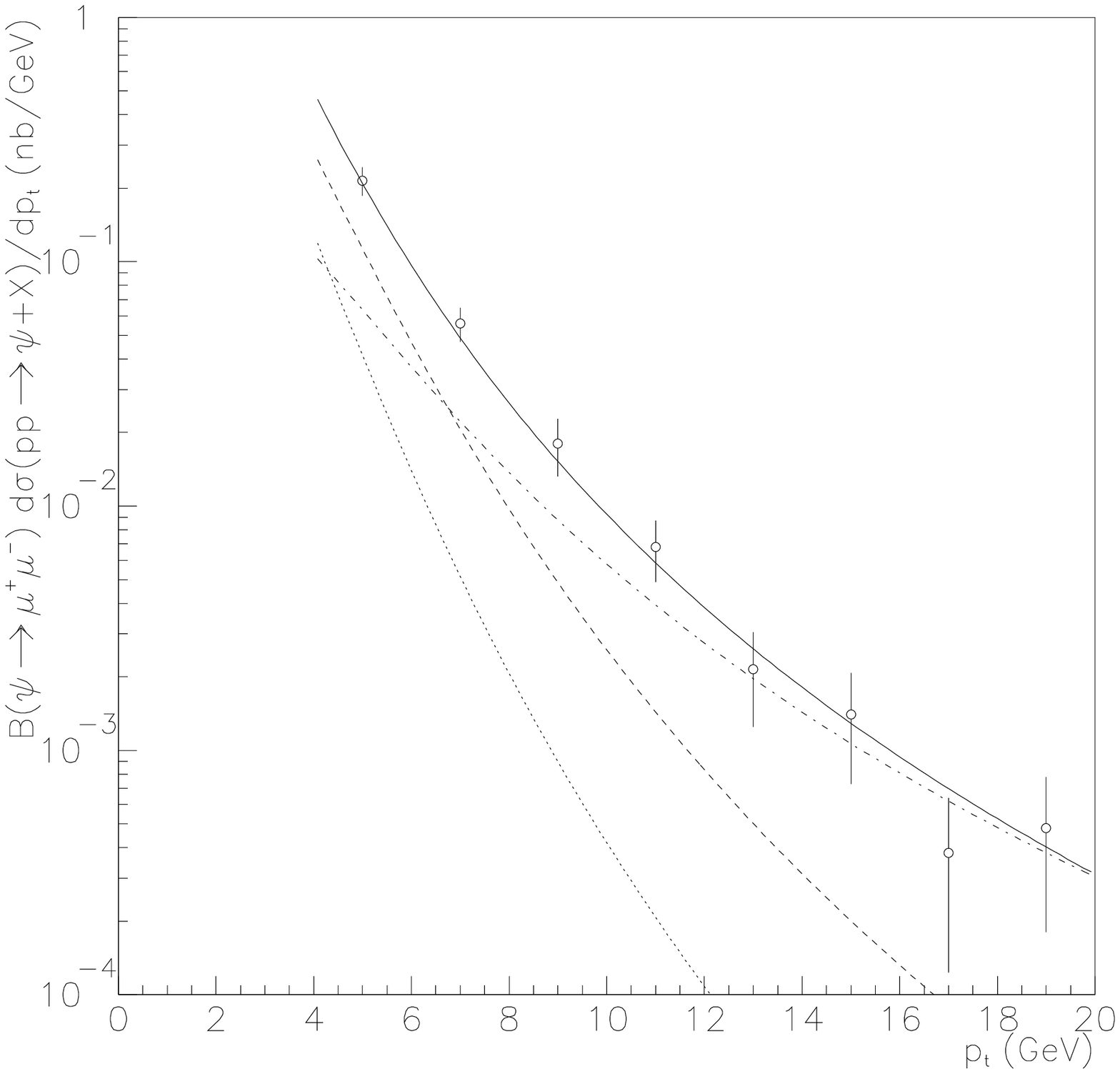,height=6.5cm,width=8cm}
}}
\caption{Two-parameter fits to the experimental Tevatron data using
CSM + COM, where initial- and final-state radiation were incorporated 
via PYTHIA generation. a) ({\em left}) $J/\psi$ and b) ({\em right})
$\psi'$ production. 
The different contributions are shown separately i) dotted line: CSM, 
ii) dashed line: $^1S_0^{(8)}$+$^3P_J^{(8)}$, iii) dot-dashed line: 
$^3S_1^{(8)}$, iv) solid line: all contributions.}
\end{figure}

\begin{table*}[hbt]
\setlength{\tabcolsep}{1.5pc}
\newlength{\digitwidth} \settowidth{\digitwidth}{\rm 0}
\caption{Colour-octet matrix elements (in units of GeV$^3$) from the 
best ${\chi}^2$ fit to Tevatron data on prompt $J/\psi$ production
for different parton distribution functions. The error bars are
statistical only. For comparison we quote the values given in Ref. [17]: 
$(6.6{\pm}2.1){\times}10^{-3}$ and $(2.2{\pm}0.5){\times}10^{-2}$  
respectively.}

\begin{center}
\begin{tabular}{lcc}    \hline
matrix element:  & $<0{\mid}O_8^{J/\psi}(^3S_1){\mid}0>$ & 
$\frac{<0{\mid}O_8^{J/\psi}(^3P_0){\mid}0>}{M_c^2}+
\frac{<0{\mid}O_8^{J/\psi}(^1S_0){\mid}0>}{3}$ \\
\hline

CTEQ2L &$(3.3{\pm}0.5){\times}10^{-3}$& $(4.8{\pm}0.7){\times}10^{-3}$ \\
MRSD0 & $(2.1{\pm}0.5){\times}10^{-3}$ & $(4.4{\pm}0.7){\times}10^{-3}$ \\
GRV 94 HO & $(3.4{\pm}0.4){\times}10^{-3}$ & $(2.0{\pm}0.4){\times}10^{-3}$ \\
\hline
\end{tabular}
\end{center}
\end{table*}

\begin{table*}[hbt]
\setlength{\tabcolsep}{1.5pc}
\caption{Colour-octet matrix elements (in units of GeV$^3$) from the 
best ${\chi}^2$ fit to Tevatron data on prompt ${\psi}'$ production
for different parton distribution functions. The error bars are
statistical only. For comparison we quote the values given in Ref. [17]: 
$(4.6{\pm}1.0){\times}10^{-3}$ and $(5.9{\pm}1.9){\times}10^{-3}$  
respectively.}

\begin{center}
\begin{tabular}{lcc}    \hline
matrix element:  & $<0{\mid}O_8^{\psi'}(^3S_1){\mid}0>$ & 
$\frac{<0{\mid}O_8^{\psi'}(^3P_0){\mid}0>}{M_c^2}+
\frac{<0{\mid}O_8^{\psi'}(^1S_0){\mid}0>}{3}$ \\
\hline

CTEQ2L &$(1.4{\pm}0.3){\times}10^{-3}$& $(1.1{\pm}0.3){\times}10^{-3}$ \\
MRSD0 & $(1.1{\pm}0.3){\times}10^{-3}$ & $(0.94{\pm}0.23){\times}10^{-3}$ \\
GRV 94 HO & $(1.3{\pm}0.2){\times}10^{-3}$ & $(0.12{\pm}0.15){\times}10^{-3}$ \\
\hline
\end{tabular}
\end{center}
\end{table*}

Notice that all the ME's are lowered systematically with respect
to the values reported in Ref. \cite{cho} for both $J/\psi$
and $\psi'$ \footnote{A caveat is in order however. In our 
generation we have not included the leading log QCD corrections into the
colour-octet cross section (causing the depletion of the fragmentation 
function of gluons into charmonium at high-$p_t$) through the interpolating 
formula (2.22) of Ref. \cite{cho0}. However, final state radiation
was incorporated in our analysis in contrast to Ref. \cite{cho0,cho}, 
softening the $p_t$ inclusive distribution as mentioned in footnote
$\#$3. The neglect of 
those higher order QCD effects should mainly affect the high-$p_t$
region, hence the $^3S_1$ matrix element which could be
somewhat underestimated.}. Basically this
is because initial-state radiation enhances the
differential cross section for $p_t>4$ GeV, so that
the non-perturbative parameters have to be re-adjusted
to restore the fit to the experimental points. 
Let us also remark that the combined $^1S_0^{(8)}$,  
$^3P_0^{(8)}$ ME's are lowered
by larger factors than those for the $^3S_1^{(8)}$ contribution. 
Consequently, the power counting rules derived from NRQCD are more 
faithfully satisfied in the overall than in Ref. \cite{cho} -
all of them should scale like $M_c^3v^7$. It is especially
significant that the numerical values obtained in our analysis for the
$^1S_0^{(8)}$+$^3P_0^{(8)}$ ME's are smaller by an order of magnitude 
with respect to those  found in Ref. \cite{cho}
in accordance with the common belief widely expressed in literature
\cite{schuler,lite}. In this sense, theoretical expectations from the 
COM might be reconciled with HERA
data on $J/\psi$ \vspace{0.1in} photoproduction \vspace{0.1in}
\cite{hera}.
\par

The contributions from the colour-octet $^3S_1^{(8)}$ and 
$^1S_0^{(8)}$ + $^3P_J^{(8)}$ states to the differential cross 
section for $J/\psi$ and $\psi'$ production are shown separately in
Fig. 2. Notice that the $^1S_0^{(8)}$ and $^3P_J^{(8)}$ components
dominate in both cases over the other production channels
at moderate $p_t$, whereas the $^3S_1^{(8)}$ channel takes over at
large  $p_t$ becoming  asymptotically responsible of
the full production \vspace{0.1in} rate.  
\par
Let us remark the especially relative smaller numerical values of the 
long-distance $^1S_0^{(8)}$, $^3P_J^{(8)}$ matrix elements regarding
the GRV PDF. This is in a qualitative agreement 
with the fact that the MRSD0 and CTEQ2L parton distribution
functions give smaller values for the gluon distribution function
at small partonic $x$ than the GRV PDF (the gluon density of the latter 
is steeper at small $x$). Hence, a gluon density with a steeper behaviour
at small $x$ steepens the $p_t$-spectrum for the $^3S_1^{(8)}$
component amounting to a larger production rate at moderate $p_t$,
decreasing the $^1S_0^{(8)}$+$^3P_J^{(8)}$ contribution from
the combined fit to the experimental inclusive \vspace{0.2in} distribution.

\section{Charmonia Production at the LHC}
Reliable calculations of inclusive production rates of charmonia
are of considerable interest at LHC experiments in many
respects \cite{atlas,cms}, as briefly outlined in the Intorduction. 
Therefore, an order-of-magnitude estimate of its production rate becomes
very valuable at the present stage of the LHC \vspace{0.1in} experiments.
\par

\begin{figure}[htb]
\centerline{\hbox{
\psfig{figure=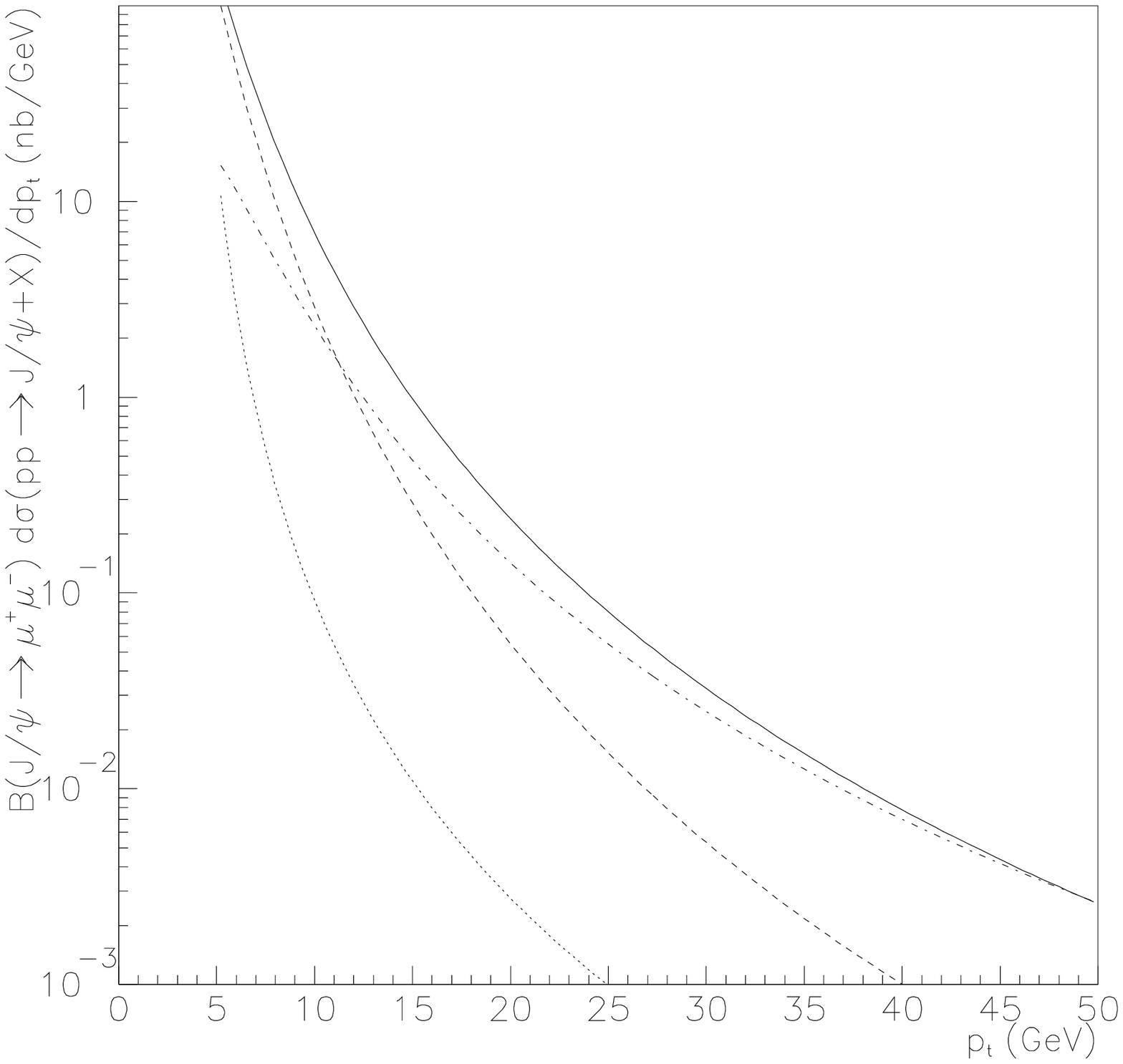,height=6.5cm,width=8.cm}
\psfig{figure=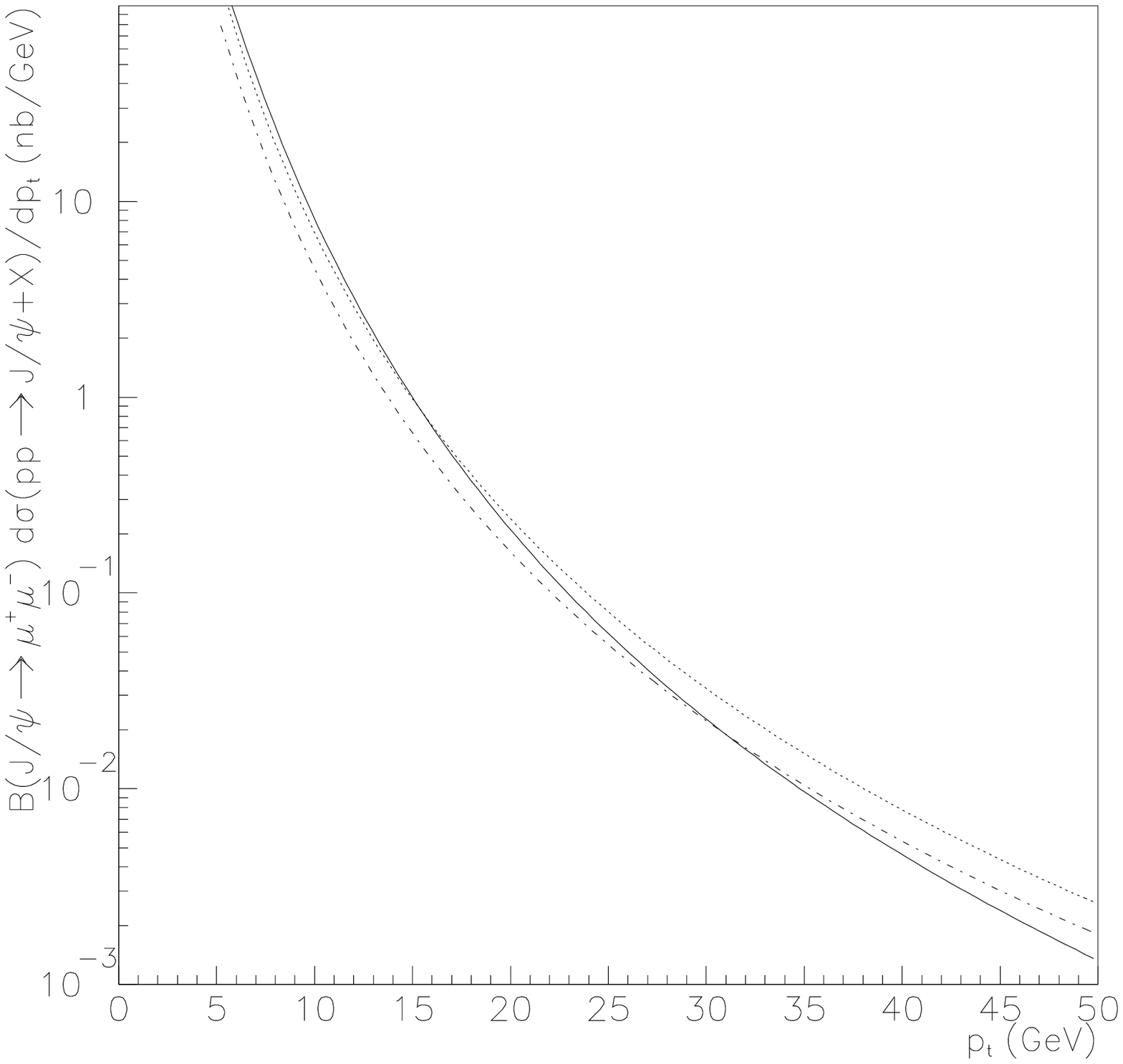,height=6.5cm,width=8.cm}
}}
\caption{({\em left}) Different contributions to $J/\psi$ hadroproduction
using the CTEQ PDF. The curves are labelled in the same way as in Fig. 2.
; ({\em right}) Our prediction for prompt $J/\psi$ 
direct production at the LHC using PYTHIA with the colour-octet matrix
elements from Table 1: a) dotted line: CTEQ 2L, b) dot-dashed line: MRSD0, 
and c) solid line: GRV HO. A rapidity cut ${\mid}y{\mid}<2.5$ was required.} 
\end{figure}

\vspace{0.1in}

\par

Keeping in mind such a variety of interests, we have generated direct
$J/\psi$'s and ${\psi}'$s in proton-proton collisions at LHC energies 
(center-of-mass energy = 14 TeV) by means of our
${\lq}{\lq}$modified" version of PYTHIA with the colour-octet ME's
as shown in Tables 1 and 2 - i.e. after normalization to Tevatron data.
The different contributions to $J/\psi$ hadroproduction
are depicted altogether in Fig. 3 for the CTEQ \vspace{0.1in} PDF.

\begin{figure}[htb]
\centerline{\psfig{figure=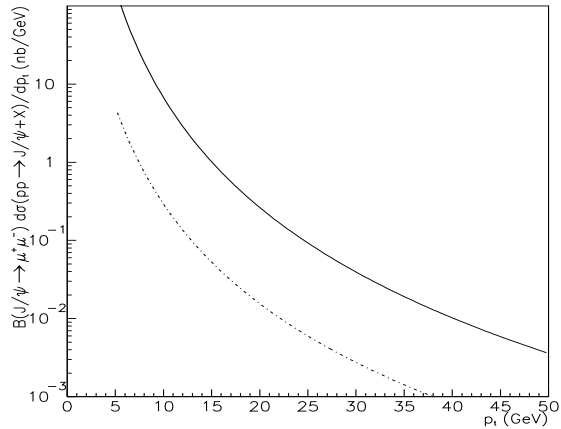,height=6.5cm,width=8.cm}}
\caption{$J/\psi$ (solid line) and ${\psi}'$ (dash-dotted line) 
direct production at the LHC according to the colour-octet ME's 
reported in Tables 1 and 2 using the CTEQ PDF. The cut 
${\mid}y{\mid}<2.5$ was required in both cases. Both curves
are multiplied by the respective branching fraction of
charmonium into muons.} 
\end{figure}

\vspace{0.1in}
\par
Figure 3 also illustrates the $p_t$ inclusive distributions of
direct prompt $J/\psi$'s at the LHC obtained for the three  PDF's employed
in our study. Comparing them with the distribution obtained by Sridhar 
\cite{sridhar}, we find our predictions standing below his theoretical curve
at large $p_t$. In fact this is not surprising since Sridhar considered only
the $^3S_1^{(8)}$ coloured intermediate state, whose ME was assumed
larger than the one used in this work. The argument ends by noticing 
that at high enough $p_t$ the dominant production comes from the 
$^3S_1^{(8)}$ channel as the combined $^1S_0^{(8)}$ + $^3P_J^{(8)}$ 
contribution falls off faster than the \vspace{0.1in} former. 
\par

In Fig. 4 we plot our prediction for direct $\psi'$ production at the LHC
using the CTEQ PDF, together with the $J/\psi$ curve. From the figure
it is clear that muon pair production from direct $J/\psi$ is larger by
more than an order of magnitude from $J/\psi$ as compared to 
${\psi}'$. It should be mentioned that our prediction for the ${\psi}'$ 
production rate differs substancially with respect to the corresponding 
theoretical curve by Sridhar \cite{sridhar} all along the $p_t$ range
- by an order of magnitude even at moderate $p_t$ in contrast to the 
$J/\psi$ \vspace{0.1in} case.
\par
\section{Conclusions}
In this paper we report on the extraction of long-distance colour-octet
matrix elements in a Monte Carlo framework from charmonia production at hadron 
colliders, already studied in our previous works \cite{bea} \cite{plb}, 
extending our analysis to the $\psi'$ resonance. We explain in more detail
the procedure followed in our investigation, also providing a code for a 
$\lq\lq$fast" simulation of charmonium in hadron collisions 
based on the colour-octet model and compatible with the
Tevatron \vspace{0.1in} results.
\par
All three possible lowest order short-distance processes contributing
to $J/\psi$ hadroproduction have been taken into account, namely, 
$g+g{\rightarrow}J/\psi+g$, $g+q{\rightarrow}J/\psi+q$ and
$q+\bar{q}{\rightarrow}J/\psi+g$, contributing at order ${\alpha}_s^3$
to charmonia production including the colour-octet mechanism. The
same set of channels have been considered for $\psi'$ 
\vspace{0.1in} hadroproduction.
\par
We have investigated higher-order effects induced by initial-state 
radiation, concluding that the overlooking of initial radiation leads 
systematically to a significant overestimate of the fit parameters. The new
colour-octet matrix elements obtained for three different PDF's 
from Tevatron data are shown in Tables 1 and 2, which have to be regarded
as reasonable estimates due to the experimental and theoretical 
uncertainties. In particular, the numerical values for the 
combined $^1S_0^{(8)}$ + $^3P_0^{(8)}$ matrix elements
are smaller by an order of magnitude than those obtained
in Ref. \cite{cho} and might resolve
the difficulties of the COM found in $J/\psi$ photoproduction at 
HERA \vspace{0.2in} \cite{hera}.
\par

\subsection*{Acknowledgments}

We thank S. Baranov, P. Eerola, N. Ellis and M. Smizanska and the ATLAS B
physics working group for useful comments and an encouraging attitude. 
Comments by E. Kovacs, T. Sj\"{o}strand and K. Sridhar are also 
acknowledged. 
\par
\newpage

\appendix 
\renewcommand{\theequation}{\thesection.\arabic{equation}}
\section*{Appendix}
\section{Code for Implementing the Colour Mechanism in
PYTHIA} \setcounter{equation}{0} \par \vspace{0.2in} 
We have modified the PYTHIA routine PYSIGH (ISUB=86) \cite{pythia} 
for the partonic process:
\[ g\ +\ g\ {\rightarrow}\ J/\psi\ +\ g \] 
including those contributions from the $^3S_1^{(8)}$, $^1S_0^{(8)}$ and 
$^3P_J^{(8)}$ coloured states according to the expressions (1) and (4). 
The ME's of the latter components are related through the heavy quark
spin symmetry at leading order in $1/M_c$: \cite{bodwin}
\[ <O_8(^3P_J)>\ =\ (2J+1)<O_8(^3P_0)> \]
Moreover, it should be stressed that 
the colour-octet ME's associated to the $^1S_0^{(8)}$ and
the $^3P_J^{(8)}$ channels cannot be 
disentangled from Tevatron experimental data since both are
degenerate in the sense that they display a similar $p_t$ 
behaviour for $p_t\ {\ge}\ 5$ GeV \cite{cho}. All three $P$-wave 
contributions altogether amount to about three times 
the $^1S_0$ contribution when setting $<O_8(^3P_0)>=M_c^2<O_8(^1S_0)>$. 
Therefore one may write the following approximate equality:
\begin{equation}
\frac{d\sigma}{dp_t}[^1S_0^{(8)}]\ 
+\sum_{J=0,1,2}\frac{d\sigma}{dp_t}[^3P_J^{(8)}]\ {\simeq}\ 
3\ {\times} \frac{1}{<O_8(^1S_0)>}\ \frac{d\sigma}{dp_t}[^1S_0^{(8)}]\ 
{\times}\ \biggl(\ \frac{<O_8(^1S_0)>}{3}+\frac{<O_8(^3P_0)>}{M_c^2}\ \biggr) 
\end{equation}
and only the $^1S_0^{(8)}$ component has to be simulated in 
\vspace{0.1in} practice.
\par
\begin{figure}[htb]
\centerline{\psfig{figure=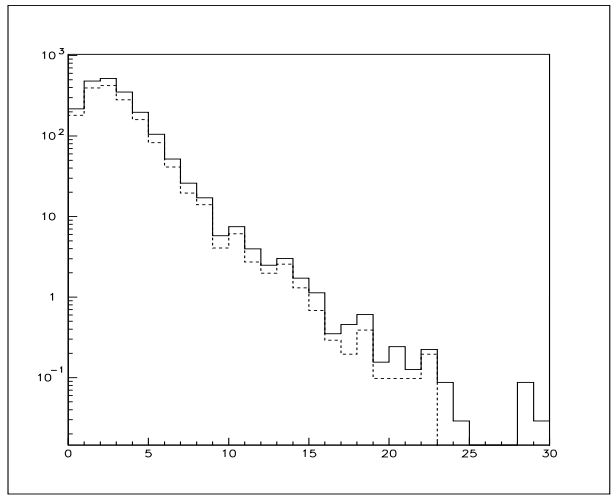,height=6.5cm,width=8.cm}}
\caption{ Histograms for 
$BF(J/\psi{\rightarrow}{\mu}^+{\mu}^-){\times}
d{\sigma}/dp_t(pp{\rightarrow}{J/\psi}X)$
obtained from PYTHIA using the CTEQ PDF at 14 TeV center-of-mass energy. 
Solid line: all contributions;
dashed line: $gg$ contribution.}
\end{figure}

\par

On the other hand, we have generated with PYTHIA an event sample 
via the gluon-quark subprocess $g+q{\rightarrow}{J/\psi}+q$
according to the expressions (2) and (5), $\lq\lq$adapting" for this
end the PYSIGH routine ISUB=112. We found that this production
channel amounts to a fraction of about $20\%$ of the total yield of
direct $J/\psi$'s, almost constant over the $p_t$ range under examination 
(see Fig. 5). Consequently, one may reduce considerably 
CPU time in the event generation by a slight change on the 
$gg{\rightarrow}{J/\psi}g$ routine, including $gq$ scattering
by means of a scaling factor. Let
us note however that all the fits and plots shown
in this work were performed using the  $gq$ event 
sample obtained from PYTHIA as explained above. The routine presented
below permits a fast generation of direct $J/\psi$'s (or ${\psi}'$ with 
some few changes in the values of the parameters and variables such as
PARP(38), SQM3, ...) 
in hadronic collisions at the \vspace{0.1in} LHC.
\par  
Thus, we have rewritten the ISUB=86 routine of PYTHIA as:

\begin{verbatim} 

      ELSEIF(ISUB.EQ.86) THEN
C...g + g -> J/Psi + g.
C
C... Color-Octet Matrix Elements for J/psi
C               (CTEQ 2L)  
C
         SQM3=(2.*1.48)**2	
         XO83S1=0.0033
         XO81S3P=0.0048
C
C   COLOR SINGLET
C
         FACQQG1=COMFAC*AS**3*(5./9.)*PARP(38)*SQRT(SQM3)*
     &    (((SH*(SH-SQM3))**2+(TH*(TH-SQM3))**2+(UH*(UH-SQM3))**2)/
     &    ((TH-SQM3)*(UH-SQM3))**2)/(SH-SQM3)**2
C
C   COLOR OCTET (3-S-1)
C
         FACQQG2=-COMFAC*AS**3*(1./18.)*3.1416/SQRT(SQM3)*
     &    (27.*(SH*TH+TH*UH+UH*SH)-19.*SQM3**2)*((TH**2+TH*UH+UH**2)**2-
     &    SQM3*(TH+UH)*(2.*TH**2+TH*UH+2.*UH**2)+SQM3**2*
     &    (TH**2+TH*UH+UH**2))*XO83S1/
     &    ((TH-SQM3)*(UH-SQM3))**2/(SH-SQM3)**2
C
C   COLOR OCTET (1-S-0)/3 + (3-P-0) !!!
C
         FACQQG3=1000.*COMFAC*(5.*3.1416*AS**3/(4.*SQRT(SQM3)))/SH*
     &    (SH**2*(SH-SQM3)**2+SH*TH*UH*(SQM3-2.*SH)+(TH*UH)**2)/(TH*UH)*
     &    ((SH**2-SQM3*SH+SQM3**2)**2-
     &    TH*UH*(2.*TH**2+3.*TH*UH+2.*UH**2))*XO81S3P/
     &    ((TH-SQM3)*(UH-SQM3))**2/(SH-SQM3)**2
          FACQQG3=FACQQG3/1000.
C
C ... Scaling factor for g + q -> J/psi + q
C
         SCALGQ=1.25
C
         IF(KFAC(1,21)*KFAC(2,21).NE.0) THEN
            NCHN=NCHN+1
            ISIG(NCHN,1)=21
            ISIG(NCHN,2)=21
            ISIG(NCHN,3)=1
            SIGH(NCHN)=FACQQG1
            NCHN=NCHN+1
            ISIG(NCHN,1)=21
            ISIG(NCHN,2)=21
            ISIG(NCHN,3)=1
            SIGH(NCHN)=FACQQG2*SCALGQ
            NCHN=NCHN+1
            ISIG(NCHN,1)=21
            ISIG(NCHN,2)=21
            ISIG(NCHN,3)=1
            SIGH(NCHN)=3.*FACQQG3*SCALGQ
         ENDIF
\end{verbatim}
\vspace{0.1in}
\par
The following variables, originally not present in PYTHIA, have
been specifically introduced in our routine:
\begin{itemize}
\item XO83S1: colour-octet ME associated to the $^3S_1$ component
\item XO81S3P: linear combination $<O_8(^1S_0)>/3$+$<O_8(^3P_0)>/M_c^2$
\item SCALGQ: scaling factor taking into account $gq$ production
\end{itemize}
\vspace{0.1in}
\par

A lower cut-off was used throughout in the event generation
($p_{tmin}=1$ GeV by default in PYTHIA for processes singular at 
$p_t{\rightarrow}0$) thereby avoiding the problematic low-$p_t$ region 
for charmonia hadroproduction. The effect of such a cut on the experimentally 
observed $p_t$ range ($p_t>4$ GeV) is rather small, representing a slight 
underestimation of the production rate in the event generation with 
radiation on as a consequence of the ISR effect \vspace{0.1in} \cite{break}.
\par
Lastly, let us point out that the theoretical curves appearing in all our
plots were obtained from histograms (filled with the corresponding
PYTHIA ouput data) fitted by means of the four-parameter $p_t$ function:
\begin{equation}
F[{\alpha}_1,{\alpha}_2,{\alpha}_3,{\alpha}_4;p_t]\ =\ 
{\alpha}_1\ \frac{p_t^{{\alpha}_2}}{({\alpha}_3+p_t^2)^{{\alpha}_4}}
\end{equation}
providing an excellent fit in the whole $p_t$ range \vspace{0.1in} examined. 
\par
\end{document}